\global\def\draftcontrol{0}

   \def\versionno{ n4rn}

\catcode`\@=11

\expandafter\ifx\csname draftcontrol\endcsname\relax\global\def\draftcontrol{0}
\fi

{\count255=\time\divide\count255 by 60
\xdef\hourmin{\number\count255}
\multiply\count255 by-60\advance\count255 by\time
\xdef\hourmin{\hourmin:\ifnum\count255<10 0\fi\the\count255}}
\def\draftdate{\number\month/\number\day/\number\year\ \ \ \hourmin }

\newcommand\makepapertitle{\par
  \begingroup
    \renewcommand\thefootnote{\@fnsymbol\c@footnote}%
    \def\@makefnmark{\rlap{\@textsuperscript{\normalfont\@thefnmark}}}%
    \long\def\@makefntext##1{\parindent 1em\noindent
            \hb@xt@1.8em{%
                \hss\@textsuperscript{\normalfont\@thefnmark}}##1}%
     \newpage
     \global\@topnum\z@   
     \@makepapertitle
     \thispagestyle{empty}\@thanks
  \endgroup
  \setcounter{footnote}{0}%
  \global\let\thanks\relax
  \global\let\makepapertitle\relax
  \global\let\@makepapertitle\relax
  \global\let\@thanks\@empty
  \global\let\@author\@empty
  \global\let\@date\@empty
  \global\let\@title\@empty
  \global\let\title\relax
  \global\let\author\relax
  \global\let\date\relax
  \global\let\and\relax
  \def\version{\let\version\@version\@gobble}
}
\def\@makepapertitle{%
  \newpage
   \ifnum\draftcontrol=1 {}
   \version\versionno
   \vskip 3em%
   \else
   \hfill\hbox to 3cm {\parbox{4cm}{\@pubnum}\hss}%
   \vskip 3em%
   \fi
   \begin{center}%
   \let \footnote \thanks
     {\LARGE {\@title}}%
     \vskip 1.5em%
     {\normalsize
       \lineskip .5em%
       \begin{tabular}[t]{c}%
         \@author
       \end{tabular}\par}%
     \vskip 1.5em%
     {\@bstract}%
     \end{center}%
     \vskip 1.5em
     \@date%
   \par
}

\gdef\@pubnum{}
\def\pubnum#1{%
  \gdef\@pubnum{#1}}

\gdef\@bstract{}
\def\Abstract#1{%
  \gdef\@bstract{%
   \parbox{\textwidth-0pc}{%
   \centerline{\bf Abstract}\penalty1000%
\kern.2cm%
\noindent
\renewcommand\baselinestretch{1.0}%
{#1}}}
}

\def\ps@paper{\let\@mkboth\@gobbletwo%
     \ifnum\draftcontrol=1
    \def\@oddfoot{\hbox to \textwidth{\tiny \versionno \hfil\tiny\draftdate}%
    \hskip -\textwidth \hbox to \textwidth{\hfil\rm\thepage\hfil}}%
     \else\def\@oddfoot{\hbox to \textwidth{\hfil\rm\thepage\hfil}}
     \fi
     \let\@evenfoot\@oddfoot
}

\def\body{\clearpage
          \pagestyle{paper}
    }

\def\@version#1{\ifnum\draftcontrol=1
\typeout{}\typeout{#1}\typeout{}
\vskip3mm\centerline{\hbox{\fbox{\normalsize{\tt DRAFT -- #1 -- }
                   {\draftdate}}}}\vskip3mm
\fi}
\let\version\@version
\long\def\eqlabel#1{\ifnum\draftcontrol=1
                    \tag@false  
                    \tag*{(\theequation) \hbox to -0.2cm{\hspace{0cm}\small{#1}\hss}}
                    \refstepcounter{equation}
                    \edef\@currentlabel{\theequation}
                    \ltx@label{#1}          
                    \else
                    \label{#1}
                    \fi
                    }
\let\st@bibitem\@bibitem
\let\st@lbibitem\@lbibitem
\ifnum\draftcontrol=1
  \def\@bibitem#1{%
    \st@bibitem{#1}\a@@label{#1}\ignorespaces}
  \def\@lbibitem[#1]#2{%
    \st@lbibitem[#1]{#2}\a@@label{#2}\ignorespaces}
  \def\a@@label#1{%
    \gdef\a@lab{\smash{\normalfont\small#1}}
    \ifvmode
      \if@inlabel
        \global\setbox\@labels\hbox{%
          \llap{\a@lab\let\a@lab\relax
                \kern\@totalleftmargin\kern\marginparsep}%
          \box\@labels}%
      \fi
    \fi}
\fi

\documentclass[12pt,letterpaper]{article}

\usepackage{amsmath,amssymb,array,calc,epsfig,rotating,bm}
\usepackage[sort]{cite}
\usepackage{graphicx}
\usepackage{psfrag,verbatim}
\usepackage{xcolor}
\usepackage{hyperref}

\ifnum\draftcontrol=1
\fi

\renewcommand\baselinestretch{1.25}
\renewcommand\section{\@startsection {section}{1}{\z@}%
                                   {-3.5ex \@plus -1ex \@minus -.2ex}%
                                   {2.3ex \@plus.2ex}%
                                   {\normalfont\large\bfseries}}
\renewcommand\subsection{\@startsection{subsection}{2}{\z@}%
                                   {-3.25ex\@plus -1ex \@minus -.2ex}%
                                   {1.5ex \@plus .2ex}%
                                   {\normalfont\normalsize\bfseries}}
\renewcommand\subsubsection{\@startsection{subsubsection}{3}{\z@}%
                                   {-3.25ex\@plus -1ex \@minus -.2ex}%
                                   {1.5ex \@plus .2ex}%
                                   {\normalfont\normalsize\it}}
\renewcommand\paragraph{\@startsection{paragraph}{4}{\z@}%
                                   {-3.25ex\@plus -1ex \@minus -.2ex}%
                                   {1.5ex \@plus .2ex}%
                                   {\normalfont\normalsize\bf}}

\numberwithin{equation}{section}

\def\revise#1       {\raisebox{-0em}{\rule{3pt}{1em}}%
                     \marginpar{\raisebox{.5em}{\vrule width3pt\
                     \vrule width0pt height 0pt depth0.5em
                     \hbox to 0cm{\hspace{0cm}{%
                     \parbox[t]{4em}{\raggedright\footnotesize{#1}}}\hss}}}}

\newcommand\nxt[1]  {\\\fnxt#1}
\newcommand{\ie}{{\it i.e.,}\ }
\newcommand{\eg}{{\it e.g.,}\ }

\def\cale         {{\cal E}}

\def\calm         {{\cal M}}
\def\caln         {{\cal N}}
\def\calo         {{\cal O}}

\def\cals         {{\cal S}}

\def\del          {\partial}

\def\sqr#1#2{{\vcenter{\vbox{\hrule height.#2pt
 \hbox{\vrule width.#2pt height#1pt \kern#1pt
 \vrule width.#2pt}\hrule height.#2pt}}}}

\def\aa1{\phi}
\def\cc1{\psi}

\catcode`\@=12

\begin{document}


\title{\bf The ordered phase of charged $\caln=4$ SYM plasma}

\date{January 3, 2025}

\author{
Alex Buchel\\[0.4cm]
\it Department of Physics and Astronomy\\ 
\it University of Western Ontario\\
\it London, Ontario N6A 5B7, Canada\\
\it Perimeter Institute for Theoretical Physics\\
\it Waterloo, Ontario N2J 2W9, Canada\\
}

\Abstract{Recently is has been shown \cite{Gladden:2024ssb} that strongly
coupled ${\cal N}=4$ supersymmetric Yang-Mills plasma with diagonal
$U(1)$ $R$-charge chemical potential is unstable at $\frac{\mu}{2\pi
T}=\sqrt{2}$. We construct a new homogeneous and isotropic phase of
this plasma that dominates in the microcanonical ensemble (but not in
the grand canonical one).  The new phase extends to arbitrary high
temperatures, \ie as $T\gg \mu$, and is characterized by an
expectation value of a dimension-2 operator, with ${\cal O}_2\propto
T^2$ --- it is the {\it ordered conformal phase}.  In the limit
$\frac\mu T\to 0$, this ordered phase has vanishing energy density
$\frac{{\cal E}}{T^4}\propto \frac{\mu^2}{T^2}$ and is a low entropy
density state $\frac{{\cal S}}{T^3}\propto \frac{\mu^2}{T^2}$.
}

\makepapertitle

\body

\version\versionno
\tableofcontents

\section{Introduction and summary}\label{intro}

Holographic duality \cite{Maldacena:1997re} establishes the equivalence between
$\caln=4$ $SU(N_c)$ supersymmetric Yang-Mills (SYM) theory and 
type IIB string theory in $AdS_5\times S^5$. The dual picture is useful
when the string theory is in the two-derivative supergravity approximation,
which corresponds to planar limit on the SYM side, and the limit
of infinitely large 't Hooft coupling constant. 
We restrict our considerations to this supergravity regime.

$\caln=4$ SYM has $SU(4)$ $R$-symmetry, and its charged plasma with a
diagonal $U(1)$ chemical potential, \ie the same chemical potential for
all $U(1)$ factors of the maximal Abelian subalgebra $U(1)^3\subset SU(4)$,
has a dual gravitational description as
a Reissner-Nordstrom (RN) black hole in asymptotically $AdS_5$ space-time.
The Gibbs free energy density of this phase is given by\footnote{We use the normalization
of the chemical potential as in \cite{Gladden:2024ssb}.} 
\begin{equation}
\Omega =-\frac{c}{2\pi^2}\left(\alpha^4+\frac 12\alpha^2 \mu^2\right)\,,\qquad
\frac{T}{\mu}=\frac{4\alpha^2-\mu^2}{4\pi\alpha\mu}\,,
\eqlabel{omsysi}
\end{equation}
where $T$ and $\mu$ are the temperature and the chemical potential correspondingly;
$c=\frac{N_c^2}{4}$ is the central charge of the SYM, and $\alpha$ is an arbitrary auxiliary
scale\footnote{This scale can be eliminated in favor  of $\frac T\mu$, but then the thermodynamic
formulas look unnecessarily complicated.}.

Note that the phase \eqref{omsysi} has an extremal limit,
\ie  as $\frac T\mu\to 0$, with the finite entropy density at the extremality.
The finite entropy density at zero temperature  is highly unusual in ordinary physical systems,
and generated a lot of research activity advocating strong gravitational quantum corrections
for near-extremal horizons \cite{Turiaci:2023wrh}. The situation might have a more mundane
resolution if the near-extremal black holes are considered in string theory, rather than in
models of gravity. Embedding RN black holes in top-down holographic examples 
opens possibilities for classical instabilities due to condensation of 
supergravity modes, associated with such embedding. A common example is the
presence of charged (under the black hole gauge field) modes of higher-dimensional
supergravity, realizing the stringy holographic superconductor \cite{Gubser:2009qm,Buchel:2024phy}.
Such a mechanism can not operate in case of strongly coupled charged $\caln=4$ plasma,
as the latter lacks appropriate charged matter.
Nonetheless, extending the arguments of \cite{Buchel:2005nt}, it was shown
in\footnote{See \cite{Harmark:1999xt} for related earlier work.}
\cite{Gladden:2024ssb} that RN black holes in $AdS_5\times S^5$ have
instabilities involving a combination of off-diagonal gauge fields and neutral scalars,
when interpreted within STU consistent truncation \cite{Behrndt:1998jd,Cvetic:1999xp,Cvetic:2000nc} of type IIB supergravity.
The instability occurs for
$\frac{\mu}{2\pi T}>\sqrt{2}$. 
From the dual gauge theory perspective the instabilities are in the hydrodynamic sound
channel\footnote{Related instabilities were studied earlier in
\cite{Buchel:2010gd,Buchel:2010wk,Buchel:2011ra}.}.

In this paper we propose that the end point of the instability identified in
\cite{Gladden:2024ssb} is the new {\it ordered phase} of the charged $\caln=4$ SYM plasma.
We now summarize the properties of this new phase: 
\begin{itemize}
\item We find that $\caln=4$ SYM plasma with a diagonal $U(1)$ $R$-symmetry chemical potential
has a  phase with a nonzero expectation value of the dimension-2 operator $\calo_2$, representing
the holographic dual of the bulk scalar mode $g\equiv X^1=X^2=\frac{1}{\sqrt{X^3}}$, where
$X^a$ are the standard scalars of the STU model \cite{Behrndt:1998jd}. 
\begin{figure}[ht]
\begin{center}
\psfrag{t}[cc][][1.0][0]{{$\ln(\mu/T)$}}
\psfrag{p}[cc][][1.0][0]{{$\ln(T/\mu-T/\mu|_{crit})$}}
\psfrag{g}[tt][][1.0][0]{{$\ln(\hat\calo_2/T^2)$}}
\psfrag{o}[bb][][1.0][0]{{$\ln(\hat\calo_2)$}}
\includegraphics[width=3in]{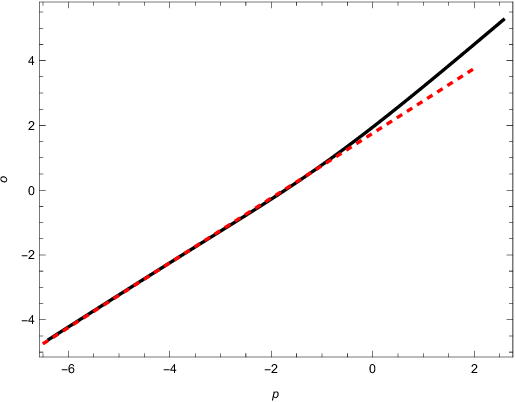}
\includegraphics[width=3in]{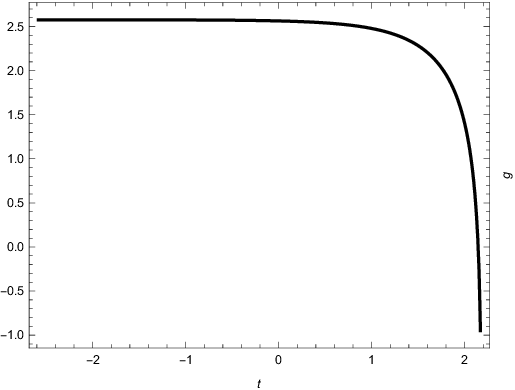}
\end{center}
  \caption{$\caln=4$ SYM with a chemical potential for a diagonal $U(1)$ of the $R$-symmetry
  has an ordered phase, characterized by the thermal expectation value of a dimension-2 operator
  $\calo_2$. The order phase exists only for $T>T_{crit}$ \eqref{tcrit}. The left panel:
  the expectation value $\calo_2$ close to criticality; the right panel: $\calo_2$ in the high temperature regime;
  see \eqref{vevo}.
  The slope of the red dashed line is 1.
} \label{fc}
\end{figure}
This phase exists
only above a critical temperature,
\begin{equation}
T>T_{crit}\equiv \frac{\mu}{2\pi \sqrt{2}}\,,
\eqlabel{tcrit}
\end{equation}
furthermore, it appears to extend to arbitrary high temperatures --- it is the {\it ordered conformal phase}\footnote{Charge neutral
conformal order was recently studied in \cite{Chai:2020zgq,Buchel:2020xdk,Buchel:2020jfs,Buchel:2020thm,Chai:2021tpt,Chaudhuri:2021dsq,Buchel:2022zxl,Chai:2021djc}.}. 
In fig.\ref{fc} we present the temperature dependence of the thermal expectation value of $\calo_2$ close to
criticality (the left panel), and in the high temperature regime $T\gg \mu$ (the right panel). The slope of the red
dashed line (the left panel) is 1. Notice that
\begin{equation}
\calo_2\propto \begin{cases}
(T-T_{crit})^1\,,\ &{\rm as}\ (T-T_{crit})\ll T_{crit}\,,\\
T^2\,,\ &{\rm as}\ \frac{T}{\mu}\gg 1\,.
\end{cases}
\eqlabel{vevo}
\end{equation}
\item The $\caln=4$ SYM ordered phase is exotic \cite{Buchel:2009ge,Buchel:2017map}: the condensate of a neutral operator
$\calo_2$ develops in the high-temperature, rather than in the low-temperature regime.

\begin{figure}[ht]
\begin{center}
\psfrag{t}[cc][][1.0][0]{{$\ln(\mu/T)$}}
\psfrag{e}[bb][][1.0][0]{{$\ln(\hat\cale/T^4)$}}
\psfrag{s}[tt][][1.0][0]{{$\ln(\hat\cals/T^3)$}}
\includegraphics[width=3in]{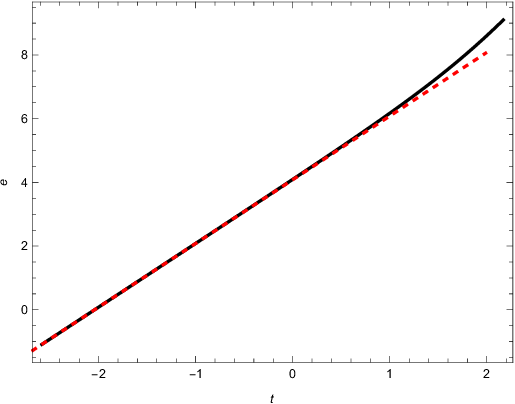}
\includegraphics[width=3in]{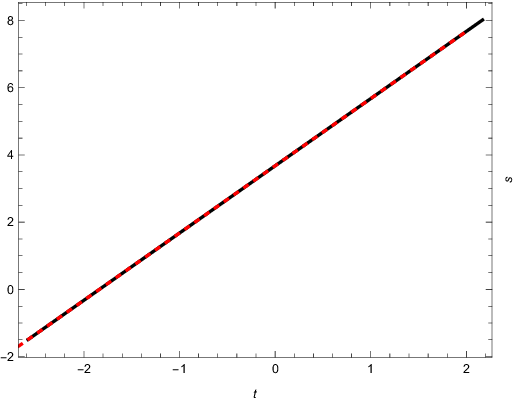}
\end{center}
  \caption{The energy density $\cale$ (the left panel)
  and the entropy density $\cals$ (the right panel) of the plasma ordered phase are small as $T\gg \mu$.
  The slopes of the red dashed lines are 2,
  resulting in the scaling relations \eqref{esscaling}.
} \label{fb}
\end{figure}

\item Although the ordered phase is 'hot', is has vanishing energy density and the entropy
in the limit $T\gg \mu$. In fig.~\ref{fb} we present the energy density $\cale$
(the left panel) and the entropy density $\cals$ of the plasma ordered phase
(the right panel). The $\ \hat{}\ $
in these thermodynamic quantities removes the central charge dependence,
see \eqref{thermo}. The dashed red lines identify the high-temperature scalings:
\begin{equation}
\frac{\hat\cale}{T^4}\propto \left(\frac{\mu}{T}\right)^2\,,\qquad \frac{\hat\cals}{T^3}\propto \left(\frac{\mu}{T}\right)^2\qquad {\rm as}\qquad \frac T\mu\to \infty\,.
\eqlabel{esscaling}
\end{equation}

\begin{figure}[ht]
\begin{center}
\psfrag{z}[tt][][1.0][0]{{$T/\mu$}}
\psfrag{t}[tt][][1.0][0]{{$\ln( T/\mu-T/\mu|_{crit})$}}
\psfrag{x}[bb][][1.0][0]{{$\hat\Omega/\mu^4$}}
\psfrag{y}[tt][][1.0][0]{{$\ln (\Delta\hat\Omega/\mu^4)$}}
\includegraphics[width=3in]{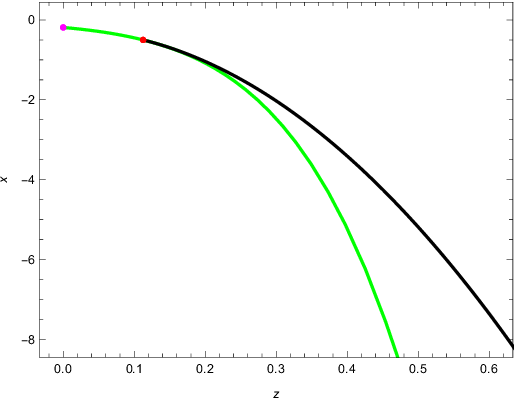}
\includegraphics[width=3in]{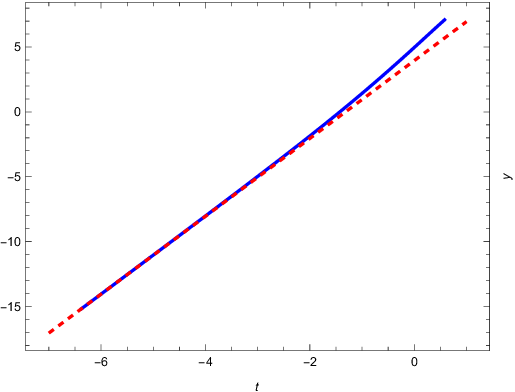}
\end{center}
  \caption{The left panel: the Gibbs free energy density of the disordered phase (the green curve)
  and the ordered phase (the black curve) of the charged $\caln=4$ SYM plasma. The red dot
  indicates the critical temperature \eqref{tcrit}; the magenta dot represents the extremal RN black hole.
  The right panel: the Gibbs free energy density difference between the ordered and the disordered and  phases close to
  criticality. The slope of the red dashed line is 3.
} \label{fa}
\end{figure}

\item  In the canonical ensemble the ordered phase is subdominant. In fig.\ref{fa} (the left panel) we compare the Gibbs free energy density $\Omega$ of
the ordered phase (the black curve) and the
disordered phase  --- with the RN black hole gravitational dual \eqref{omsysi} --- (the green curve)  .
The magenta dot represents the extremal $T=0$ limit, and the red dot represents the onset of the
hydrodynamic instability identified in \cite{Gladden:2024ssb}. In the right panel we extract the
near-critical scaling of the Gibbs free energy density difference between the phases (the solid blue curve):
\begin{equation}
\frac{\Delta\hat\Omega}{\mu^4}\equiv \frac{\hat\Omega_{ordered}-\hat\Omega_{disordered}}{\mu^4}\propto (T-T_{crit})^3\,,\qquad (T-T_{crit})\ll T_{crit}\,.
\eqlabel{dom}
\end{equation}

\begin{figure}[ht]
\begin{center}
\psfrag{x}[cc][][1.0][0]{{$\hat\cale/(\hat\rho)^{4/3}$}}
\psfrag{y}[bb][][1.0][0]{{$\hat\cals/\hat\rho$}}
\psfrag{z}[tt][][1.0][0]{{$\hat\cals/\hat\rho$}}
\includegraphics[width=3in]{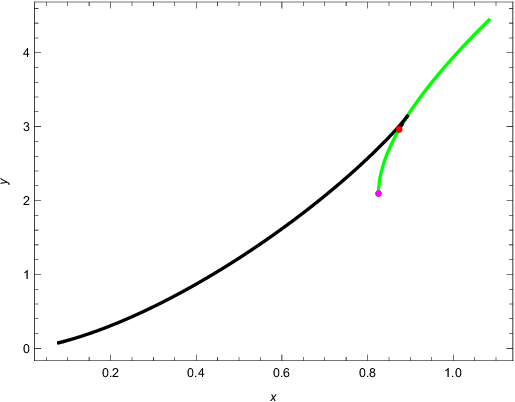}
\includegraphics[width=3in]{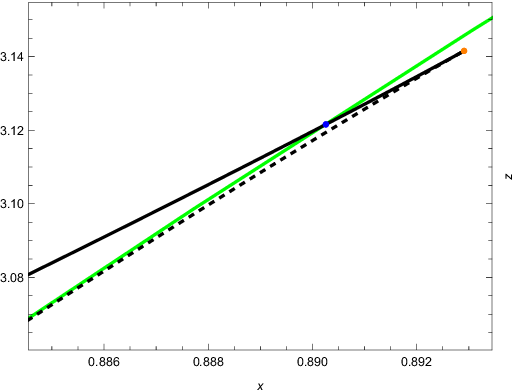}
\end{center}
  \caption{The ordered phase (the solid black curves) is favored over the disordered phase
  (the solid green curves) in the microcanonical ensemble for energy densities $\cale\le \cale_{blue}$
  \eqref{eblue}. The ordered phase is likely to extend to arbitrary low energy densities. 
} \label{figure2}
\end{figure}

\item In the microcanonical ensemble, the ordered phase is a dominant one. In fig.~\ref{figure2}
we compare the entropy density $\cals$ of the ordered (the black curves) and the disordered (the green curves)
phases at fixed charge density $\rho$, and as a function of the energy density $\cale$.
The disordered phase (the left panel) ends at the magenta point --- representing the extremal RN black hole
with the finite charge, the entropy, and the energy densities. The ordered phase originates from
the red dot (the onset of the hydrodynamic instability of the $\caln=4$ SYM disordered phase), and is always
more entropic than the disordered phase for $\hat\cale< \hat\cale_{crit}$,
\begin{equation}
\frac{\hat\cale}{(\hat\rho)^{4/3}}\bigg|_{crit}= \left(\frac 23\right)^{1/3}\,.
\eqlabel{ecrit}
\end{equation}
Actually, the ordered phase is even favored at slightly larger energy densities, \ie  up to (the blue dot of the right panel)
\begin{equation}
\frac{\hat\cale}{(\hat\rho)^{4/3}}\bigg|_{blue}=1.0190(9)\  \frac{\hat\cale}{(\hat\rho)^{4/3}}\bigg|_{crit}\,.
\eqlabel{eblue}
\end{equation}
The ordered phase appears to exist to arbitrary low energies, but terminates at the orange dot (the right panel),
\begin{equation}
\frac{\hat\cale}{(\hat\rho)^{4/3}}\bigg|_{orange}=1.0221(2)\  \frac{\hat\cale}{(\hat\rho)^{4/3}}\bigg|_{crit}\,.
\eqlabel{eorange}
\end{equation}
In the range $\hat\cale_{crit}<\hat\cale<\hat\cale_{orange}$, there appears to be two ``distinct'' ordered phases:
represented by a dashed and a solid black lines in the right panel. We leave exploration of this criticality
to the future, and only mention here that the corresponding  feature is absent in the grand canonical ensemble
discussed above. It is tempting to conjecture that the phase transition from the
disordered to the ordered phases for $\cale <\cale_{crit}$ occurs via the
hydrodynamic instability identified in \cite{Gladden:2024ssb}; but is driven by bubble nucleation
in the energy range $\hat\cale_{crit}<\hat\cale<\hat\cale_{blue}$.
\end{itemize}

In section \ref{tech} we present the technical details, sufficient to reproduced the
results reported. We conclude in section \ref{conc}.

\section{Technical details}\label{tech}

\subsection{Effective action and equations of motion}

The starting point is the STU consistent truncation of type IIB supergravity \cite{Behrndt:1998jd}: 
\begin{equation}
\begin{split}
S_{eff}=\frac{1}{2\kappa_5^2} \int_{\calm_5} \biggl(
&R-\frac 14 G_{ab}F_{\rho\sigma}^a F_{\mu\nu}^b g^{\rho\mu}g^{\sigma\nu}+
\frac{c_{abc}}{48 \sqrt{2}}\epsilon^{\mu\nu\rho\sigma\lambda}F^a_{\mu\nu}F_{\rho\sigma}^b A^c_\lambda
\\&
\qquad -G_{ab} g^{\mu\nu}\del_\mu X^a\del_\nu X^b+\sum_{a=1}^3 \frac{4}{X^a}
\biggr) \star 1\,,
\end{split}
\eqlabel{seff}
\end{equation}
where $c_{abc}$ are symmetric constants, nonzero only for distinct indices with $c_{123}=1$, $g_{\mu\nu}$ is the metric on $\calm_5$, $F^a_{\mu\nu}$ are the field strengths for the gauge fields $A^a_\mu$, $a=1\cdots 3$,
dual to conserved currents of the maximal Abelian subgroup of the $SU(4)$ $R$-symmetry of $\caln=4$ SYM. The three
real positive neutral scalar fields $X^a$  describe the deformation of $S^5$ in the uplift of $S_{eff}$ to type IIB supergravity; they are
constrained, at the level of the effective action \eqref{seff}, by
\begin{equation}
X^1 X^2 X^3=1\,.
\eqlabel{constr}
\end{equation}
The field space metric $G_{ab}$ is
\begin{equation}
G_{ab}=\frac 12 {\rm diag} \biggl((X^1)^{-2}\,,\, (X^2)^{-2}\,,\, (X^3)^{-2}\biggr)\,.
\eqlabel{const}
\end{equation}
The gravitational constant $\kappa_5$ is related to the central charge $c$ of the boundary gauge theory
as
\begin{equation}
\kappa_5^2=\frac{\pi^2}{c}=\frac{4\pi^2 }{N_c^2}\,.
\eqlabel{k5}
\end{equation}

To consider $R$-charged black holes in $S_{eff}$ \eqref{seff}, which realize the gravitational dual
to equilibrium thermal states of $\caln=4$  SYM plasma at finite $R$-symmetry chemical potentials,
we take the following ansatz:
\begin{equation}
\begin{split}
&ds_5^2=g_{\mu\nu}dx^\mu dx^\nu= -{c}_1^2\ dt^2+{c}_2^2\ d\bm{x}^2+{c}_3^2\ dr^2\,,\qquad {c}_i
={c}_i(r)\,,\\
&A^a=a_t^a(r)\ dt\,,\qquad X^1=x_1(r)\,,\qquad X^2=x_2(r)\,.
\end{split}
\eqlabel{backw1}
\end{equation}

It is straightforward to obtain equations of motion
for the fields in \eqref{backw1}:
\begin{equation}
\begin{split}
&0=(a_t^1)''+\biggl(
-\frac{2 x_1'}{x_1}-\frac{c_1'}{c_1}-\frac{c_3'}{c_3}+\frac{3 c_2'}{c_2}\biggr) (a_t^1)'\,,
\end{split}
\eqlabel{eom1}
\end{equation}
\begin{equation}
\begin{split}
&0=(a_t^2)''+\biggl(
-\frac{2 x_2'}{x_2}-\frac{c_1'}{c_1}-\frac{c_3'}{c_3}+\frac{3 c_2'}{c_2}\biggr) (a_t^2)'\,,
\end{split}
\eqlabel{eom2}
\end{equation}
\begin{equation}
\begin{split}
&0=(a_t^3)''+\biggl(
\frac{2 x_1'}{x_1}+\frac{2 x_2'}{x_2}-\frac{c_1'}{c_1}-\frac{c_3'}{c_3}+\frac{3 c_2'}{c_2}\biggr) (a_t^3)'\,,
\end{split}
\eqlabel{eom3}
\end{equation}
\begin{equation}
\begin{split}
&0=x_1''-\frac{(x_1')^2}{x_1}+\biggl(
\frac{c_1'}{c_1}-\frac{c_3'}{c_3}+\frac{3c_2'}{c_2}\biggr) x_1'+\frac{4 c_3^2}{3 x_2} (x_2^2 x_1^2-2 x_2+x_1)
+\frac{1}{6 x_1 x_2^2 c_1^2} \biggl(\\
&((a_t^3)')^2 x_1^4 x_2^4-2 x_2^2 ((a_t^1)')^2+x_1^2 ((a_t^2)')^2
\biggr)\,,
\end{split}
\eqlabel{eom4}
\end{equation}
\begin{equation}
\begin{split}
&0=x_2''-\frac{(x_2')^2}{x_2}+\biggl(
\frac{c_1'}{c_1}-\frac{c_3'}{c_3}+\frac{3c_2'}{c_2}\biggr) x_2'+\frac{4 c_3^2}{3 x_1} (x_2^2 x_1^2-2 x_1+x_2)
+\frac{1}{6 x_2 x_1^2 c_1^2} \biggl(\\
&((a_t^3)')^2 x_1^4 x_2^4-2 x_1^2 ((a_t^2)')^2+x_2^2 ((a_t^1)')^2\biggr)\,,
\end{split}
\eqlabel{eom5}
\end{equation}
\begin{equation}
\begin{split}
&0=c_1''-\frac{c_1}{c_2^2} (c_2')^2+\left(-\frac{c_3'}{c_3}+\frac{2 c_2'}{c_2}\right) c_1'-\frac{5}{24x_1^2 x_2^2 c_1}
\biggl(((a_t^3)')^2 x_1^4 x_2^4+x_2^2 ((a_t^1)')^2\\& +x_1^2 ((a^2_t)')^2\biggr)
+\frac{c_1}{6} \biggl(
\frac{(x_1')^2}{x_1^2}+\frac{x_1' x_2'}{x_1 x_2}+\frac{(x_2')^2}{x_2^2}\biggr)
-\frac{2c_1 c_3^2 (x_2^2 x_1^2+x_2+x_1)}{3x_1 x_2}\,,
\end{split}
\eqlabel{eom6}
\end{equation}
\begin{equation}
\begin{split}
&0=c_2''+\frac{(c_2')^2}{c_2}-\frac{c_3' c_2'}{c_3}+\frac{c_2}{24x_2^2 c_1^2 x_1^2} \biggl(
((a^3_t)')^2 x_1^4 x_2^4+x_2^2 ((a^1_t)')^2+x_1^2 ((a^2_t)')^2\biggr)
\\&+\frac{c_2}{6} \biggl(
\frac{(x_1')^2}{x_1^2}+\frac{x_1'x_2'}{x_1 x_2}+\frac{(x_2')^2}{x_2^2}\biggr)
-\frac{2c_3^2 c_2}{3x_1 x_2} (x_2^2 x_1^2+x_2+x_1)\,,
\end{split}
\eqlabel{eom7}
\end{equation}
\begin{equation}
\begin{split}
&0=\frac{6 (c_2')^2}{c_2^2}+\frac{6 c_2'c_1'}{c_1 c_2}-\frac{(x_2')^2}{x_2^2}-\frac{x_1'x_2'}{x_1 x_2}-\frac{(x_1')^2}{x_1^2}-4 c_3^2
\biggl(
x_1 x_2+\frac{1}{x_2}+\frac{1}{x_1}\biggr)\\&+\frac{1}{4 c_1^2} \biggl(
x_1^2 x_2^2 ((a^3_t)')^2+\frac{((a^1_t)')^2}{x_1^2}+\frac{((a^2_t)')^2}{x_2^2}
\biggr)\,.
\end{split}
\eqlabel{eomc}
\end{equation}

There is an obvious solution to the equations of motion, representing the
$\caln=4$ SYM  plasma with a diagonal chemical potential --- the disordered phase of the SYM plasma:
\begin{equation}
\begin{split}
&c_1=\frac{\alpha\sqrt{f}}{\sqrt{r}}\,,\qquad c_2=\frac{\alpha}{\sqrt{r}}\,,\qquad c_3=\frac{s}{2r\sqrt{f}}
\,,\qquad f=1-r^2\left(1+\frac{\kappa^2}{9\alpha^2}\right)+\frac{\kappa^2}{9\alpha^2} r^3\,,\\
&a_t^1=a_t^2=a_t^3=\frac{\kappa\sqrt{2}}{3} (1-r)\,,\qquad s=x_1=x_2=1\,,
\end{split}
\eqlabel{rnbh}
\end{equation}
where the radial coordinate $r\in (0,1)$, with $r\to 0_+$ being the asymptotic
$AdS_5$ boundary, and $r\to 1_-$ being a regular Schwarzschild horizon.
Parameter $\kappa$ is related to a  chemical potential $\mu$, and $\alpha$ determines the
Hawking temperature of the black hole as follows:
\begin{equation}
\mu=\frac{\kappa\sqrt{2}}{3}\,,\qquad T=\frac{18\alpha^2-\kappa^2}{18\pi\alpha}\,.
\eqlabel{tkw}
\end{equation}
Without the loss of generality we can always set $\alpha=1$, as long as we remember to represent
all the physical quantities as dimensionless ratios --- we will do so from now on.
Note that the chemical potential in the disordered phase varies as
\begin{equation}
\mu\in [0,2)\,;
\eqlabel{murange}
\end{equation}
as $\mu\to 2$ the Hawking temperature vanishes, and the black hole becomes extremal.

\subsection{Linearized fluctuations}\label{lin}

Next, we move to the discussion of the generic fluctuations about the
background \eqref{rnbh}. Introducing
\begin{equation}
\begin{split}
&f=\biggl(1-r^2\left(1+\frac{\kappa^2}{9\alpha^2}\right)+\frac{\kappa^2}{9\alpha^2} r^3\biggr)(1+\delta f)\,,\qquad s=1+\delta s\,,\\
&a_t^a=\frac{\kappa\sqrt{2}}{3} (1-r)(1+\delta v_a)\,,\qquad x_1=\exp(\delta a)\,,\qquad x_2=\exp(\delta b)\,,
\end{split}
\eqlabel{fluc}
\end{equation}
to linear order in $\delta$-fluctuations we find two identical decoupled sets:
\begin{itemize}
\item (A):
\begin{equation}
\begin{split}
&0=\delta a''+\frac{2 k^2 r^3-k^2 r^2-9 r^2-9}{(r-1) (k^2 r^2-9 r-9) r}\ \delta a'
+\frac{2 k^2 r^3+9}{(k^2 r^2-9 r-9) r^2 (r-1)}\ \delta a\\
&-\frac{4 r}{3k^2 r^2-9 r-9}\ \delta u_a'
-\frac{4r}{3(r-1) (k^2 r^2-9 r-9)}\ \delta u_a \,,
\end{split}
\eqlabel{a1}
\end{equation}
\begin{equation}
\begin{split}
&0=\delta u_a''+\frac{2}{r-1}\ \delta u_a'-\frac{3 k^2}{r-1}\ \delta a'\,;
\end{split}
\eqlabel{a2}
\end{equation}
\item (B):
\begin{equation}
\begin{split}
&0=\delta b''+\frac{2 k^2 r^3-k^2 r^2-9 r^2-9}{(r-1) (k^2 r^2-9 r-9) r}\ \delta b'
+\frac{2 k^2 r^3+9}{(k^2 r^2-9 r-9) r^2 (r-1)}\ \delta b\\
&-\frac{4 r}{3k^2 r^2-9 r-9}\ \delta u_b'
-\frac{4r}{3(r-1) (k^2 r^2-9 r-9)}\ \delta u_b \,,
\end{split}
\eqlabel{b1}
\end{equation}
\begin{equation}
\begin{split}
&0=\delta u_b''+\frac{2}{r-1}\ \delta u_b'-\frac{3 k^2}{r-1}\ \delta b'\,,
\end{split}
\eqlabel{b2}
\end{equation}
\end{itemize}
where we introduced $\delta u_a$ and $\delta u_b$ so that 
\begin{equation}
\delta v_1=\delta v_3+\frac{4}{3k^2}\ \delta u_a+\frac{2}{3k^2}\ \delta u_b\,,\qquad
\delta v_2=\delta v_3+\frac{4}{3k^2}\ \delta u_b+\frac{2}{3k^2}\ \delta u_a\,.
\eqlabel{uaub}
\end{equation}
Once the solutions to \eqref{a1}-\eqref{b2} are found, the remaining fluctuations are determined
from
\begin{equation}
\begin{split}
&0=\delta v_3''+\frac{2}{r-1}\ \delta v_3' +\frac{2(\delta a+\delta b)-\delta s'}{r-1} \,,
\end{split}
\eqlabel{fl3}
\end{equation}
\begin{equation}
\begin{split}
&0=\delta f'-\frac{4r^2}{3(k^2 r^2-9 r-9)}
\biggl(
\delta u_a'+\delta u_b'+\frac32 k^2\ \delta v_3' \biggr)+\frac{(k^2 r^3-18) \delta f+36 \delta s}
{(k^2 r^2-9 r-9) (r-1) r}\\
&-\frac{2(3 \delta v_3 k^2+2 \delta u_a+2 \delta u_b) r^2}{3(r-1) (k^2 r^2-9 r-9)}\,,
\end{split}
\eqlabel{flf}
\end{equation}
\begin{equation}
\begin{split}
&0=\delta s'\,.
\end{split}
\eqlabel{fls}
\end{equation}

We show in appendix \ref{mode3} that the fluctuation sets (A) and  (B),
associated with operators of scaling dimension $\Delta=2$ ($\delta a$ and $\delta b$) develop 
an expectation value, signaling the instability, at (see \eqref{tkw})
\begin{equation}
\kappa_{crit}=3\qquad \Longrightarrow\qquad  \mu_{crit}=\sqrt{2}\qquad \Longrightarrow \qquad
\frac{\mu}{2\pi T}\bigg|_{crit}=\sqrt{2} \,,
\eqlabel{kcr}
\end{equation}
precisely as established in \cite{Gladden:2024ssb}.

\subsection{Consistent truncation}\label{non}

In this  section we provide technical details for the construction of the
fully nonlinear solutions, associated with the onset of the linearized instabilities
identified in section \ref{lin}.
Notice that the two independent modes $\delta a$ and $\delta b$ become simultaneously
unstable in the linearized approximation\footnote{Such phenomenon was first observed in \cite{Buchel:2024phy}.} ---
thus the full phase structure of the theory can be rather involved.
Here, we restrict the discussion to a consistent truncation of \eqref{seff},
\begin{equation}
x_1\equiv x_2\equiv g\,,\qquad a_t^1\equiv a_t^2\,,
\eqlabel{ab}
\end{equation}
that captures the instability, but might not be the full story. 
Of course, there are two additional, identical to \eqref{ab}, consistent truncations : 
\begin{equation}
\biggl\{
x_1\equiv \underbrace{(x_1x_2)^{-1}}_{X^3}\,,\ a_t^1\equiv a_t^3\biggr\}\qquad {\rm or}\qquad
\biggl\{x_2\equiv \underbrace{(x_1x_2)^{-1}}_{X^3} \,,\ a_t^2\equiv a_t^3\biggr\}\,.
\eqlabel{more}
\end{equation}

Parameterizing the 5d metric  as in \eqref{rnbh},
\begin{equation}
c_1=\frac{\sqrt f}{\sqrt r}\,,\qquad c_2=\frac{1}{\sqrt r}\,,\qquad c_3=\frac{s}{2r\sqrt{f}}\,,
\eqlabel{5dmetf}
\end{equation}
the equations of motion describing black holes with scalar hair, \ie the condensate of $g$,
are given by: 
\begin{equation}
\begin{split}
&0=(a^1_t)''+\frac{2 g' (g' r-g)}{g^2} \ (a^1_t)'\,,
\end{split}
\eqlabel{fin1}
\end{equation}
\begin{equation}
\begin{split}
&0=(a^3_t)''+\frac{2 g' (g' r+2 g)}{g^2}\ (a_3^t)'\,,
\end{split}
\eqlabel{fin2}
\end{equation}
\begin{equation}
\begin{split}
&0=g''-\frac{(g')^2}{g}+\biggl(
\frac{g^4 r^2 ((a^3_t)')^2}{6f}-\frac{2g^2 s^2}{3r f}+ \frac{r^2 ((a^1_t)')^2}{3g^2 f}
-\frac{4s^2}{3g r f}+\frac1r\biggr)\ g'+\frac{g^5 r ((a^3_t)')^2}{6f}
\\&+\frac{g^3 s^2}{3r^2 f}- \frac{r ((a^1_t)')^2}{6g f}-\frac{s^2}{3r^2 f}\,,
\end{split}
\eqlabel{fin3}
\end{equation}
\begin{equation}
\begin{split}
&0=f'+\frac{1}{6g^2 r} \biggl(
-g^6 ((a^3_t)')^2 r^3+4 g^4 s^2+12 f (g')^2 r^2-2 ((a^1_t)')^2 r^3-12 g^2 f+8 g s^2
\biggr)\,,
\end{split}
\eqlabel{fin4}
\end{equation}
\begin{equation}
\begin{split}
&0=s'+\frac{2 s r (g')^2}{g^2}\,.
\end{split}
\eqlabel{fin5}
\end{equation}
They are solved subject to the following asymptotics:
\nxt in the UV, \ie as $r\to 0_+$,
\begin{equation}
\begin{split}
&f=1+F_4\ r^2+\left(\frac 13 a_1^2+\frac 16 b_1^2\right)\ r^3+\calo(r^4)\,,\qquad
s=1- g_2^2\ r^2+\frac43 g_2^3\ r^3+\calo(r^4)\,,
\end{split}
\eqlabel{uvgen1}
\end{equation}
\begin{equation}
\begin{split}
&a_t^1=\mu+a_1\ r+a_1 g_2\ r^2+\calo(r^4)\,,\qquad a_t^3=\mu+b_1\ r-2b_1 g_2\ r^2+3g_2^2 b_1\ r^3+\calo(r^4)\,,
\end{split}
\eqlabel{uvgen2}
\end{equation}
\begin{equation}
\begin{split}
&g=1+g_2\ r +\left(-\frac 13 g_2^3-\frac 14 g_2 F_4 +\frac{1}{24}a_1^2-\frac{1}{24}b_1^2\right)\ r^3+\calo(r^4)\,,
\end{split}
\eqlabel{uvgen5}
\end{equation}
specified by 
\begin{equation}
\biggl\{\
g_2\,,\, a_{1}\,,\, b_{1}\,,\, F_4\ 
\biggr\}\,,
\eqlabel{uvpar}
\end{equation}
as functions of the diagonal chemical potential $\mu$; 
\nxt in the IR, \ie as $y\equiv 1-r\to 0_+$,
\begin{equation}
\begin{split}
&g=g^h_0+\calo(y)\,,\qquad s=s^h_0+\calo(y)\,,\qquad a_t^1=a_1^h\ y+\calo(y^2)\,,\qquad a_t^3=b_1^h\ y+\calo(y^2)\,,\\
&f=-\frac{(b^h_1)^2 (g^h_0)^6-4(g^h_0)^4(s^h_0)^2-8g^h_0(s^h_0)^2+2(a^h_1)^2)}{6g^h_0}\ y+\calo(y^2)\,,
\end{split}
\eqlabel{irass}
\end{equation}
specified by 
\begin{equation}
\biggl\{\
g_0^h\,,\ s^h_0\,,\, a_1^h\,,\, b_1^h
\
\biggr\}\,,
\eqlabel{irpar}
\end{equation}
again, as functions of the diagonal chemical potential $\mu$. 

In practice, we solve equations of motion for $f,s,a^1_t,a^3_t$ and $g$ 
using the shooting method codes adopted from \cite{Aharony:2007vg}.

Once hairy black hole solutions are constructed, we can use the holographic
renormalization\footnote{In this model one should be careful
with the holographic renormalization: even thought there is no source term for the
bulk mode $g$, the counterterms for renormalization of $\Delta=2$ modes are needed
to insure the correct thermodynamics;
see \eg \cite{Buchel:2012gw}.}
to extract their thermodynamic properties:
\begin{equation}
\begin{split}
&2\pi T=\left(\frac{2(g^h_0)^2}{3}+\frac{4}{3g^h_0}\right)s^h_0-\frac{1}{6s^h_0}
\left((g^h_0)^4 (b_1^h)^2+\frac{2 (a^h_1)^2}{(g^h_0)^2}\right)\,,\qquad \hat\calo_2\equiv 2\kappa_5^2\calo_2=g_2\,,
\\
&\hat{\cals}\equiv 2\kappa_5^2 \cals =4\pi\,,\qquad  \hat\Omega\equiv 2\kappa_5^2 \Omega =\mu (2 a_1+b_1) -3 F_4 -T\hat\cals\,,\\
&\hat\cale\equiv 2\kappa_5^2 \cale = -3\hat\Omega\,,\qquad \hat\rho\equiv  2\kappa_5^2 \rho =\frac 1\mu \left(
\hat\cale-T\hat\cals-\hat\Omega\right)\,,
\end{split}
\eqlabel{thermo}
\end{equation}
where $T$ is the temperature,  $\cals$ is the entropy density, $\Omega$ is the Gibbs free energy density,
$\cale$ is the energy density, $\rho$ is the $R$-symmetry charge density, and $\calo_2$ is the thermal
expectation value of the dimension-2 operator dual to the bulk scalar $g$.

Given the solution \eqref{rnbh}, we identify
\begin{equation}
\begin{split}
&2\pi T=2\alpha-\frac{\mu^2}{2\alpha}\,,\qquad \hat\cals=4\pi\alpha^3\,,\qquad \hat\Omega=-\alpha^4
-\frac 12\alpha^2\mu^2\,,\\
&\hat\cale=3\alpha^4+\frac 32\alpha^2\mu^2\,,\qquad \hat\rho=3\mu\alpha^2\,, \qquad \hat\calo_2=0\,,
\end{split}
\eqlabel{symthermo}
\end{equation}
where we restored $\alpha$. From \eqref{symthermo} we can verify the fundamental thermodynamic relation
\begin{equation}
d\hat\Omega=-\hat\cals\ dT-\hat\rho\ d\mu\,.
\eqlabel{stlaw}
\end{equation}

\begin{figure}[ht]
\begin{center}
\psfrag{t}[tt][][1.0][0]{{$T/\mu$}}
\psfrag{z}[tt][][1.0][0]{{$1/\hat\cals\cdot \del\hat\Omega/\del T+1$}}
\psfrag{y}[tt][][1.0][0]{{$1/\hat\cals\cdot \del\hat\Omega/\del T+1$}}
\includegraphics[width=3in]{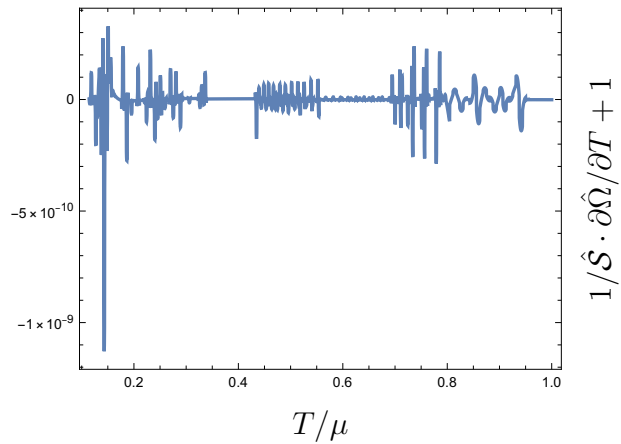}
\end{center}
  \caption{Numerical test of the first law of thermodynamics \eqref{1stlaw}
  for the scalarized charged black holes.
} \label{figure3}
\end{figure}

In the  phase of the model with scalar hair black holes, \ie the ordered phase of the charged SYM plasma, \eqref{stlaw} provides an important check on numerics.
In fig.~\ref{figure3} we present the check of the first law of thermodynamics, namely
\begin{equation}
0=\frac{\mu^3}{\hat\cals}\cdot \left(\frac{\del \frac{\hat\Omega}{\mu^4}}{\del \frac T\mu}\right)\bigg|_{\mu={\rm const}} +1\,.
\eqlabel{1stlaw}
\end{equation}

\section{Conclusion}\label{conc}

In this paper we identities a novel phase of $\caln=4$ SYM plasma, charged under the diagonal $U(1)$ $R$-symmetry.
This new phase is a {\it conformal order phase} --- it is characterize by the expectation value of a charge-neutral dimension-2 operator,
and extends to arbitrary high temperatures. The described phenomenon occurs in other holographic models as well\footnote{In preparation.}. 
Of course, much is left for future exploration --- it would be extremely interesting to understand the dynamics of the phase transition
between the ordered and the disordered phases of the charged $\caln=4$ SYM plasma. It is also possible
that additional phases of the model exist, given that two distinct operators become simultaneously unstable
in the linearized approximation at the critical temperature \eqref{tcrit}.

The existence of ordered phases might be a welcome news for the extremal black holes in string theory:
it simplifies the conundrum of the extremal horizons, namely, it is possible that they
always become classically unstable well before the quantum effects kick in.

\section*{Acknowledgments}
Research at Perimeter
Institute is supported by the Government of Canada through Industry
Canada and by the Province of Ontario through the Ministry of
Research \& Innovation. This work was further supported by
NSERC through the Discovery Grants program.

\appendix
\section{Onset of instability of $\{\delta a\,,\ \delta u_a\}$ fluctuations}\label{mode3}

The easiest way to identify the onset of an instability for a gravitational mode is to turn on
its source term; the instability is then signalled by the divergence of its normalizable
component\footnote{See for example section 5 of \cite{Buchel:2019pjb}.}.

\begin{figure}[ht]
\begin{center}
\psfrag{z}[tt][][1.0][0]{{$\kappa$}}
\psfrag{x}[bb][][1.0][0]{{$1/a_{2,0}$}}
\psfrag{y}[tt][][1.0][0]{{$1/u_{1,0}$}}
\includegraphics[width=3in]{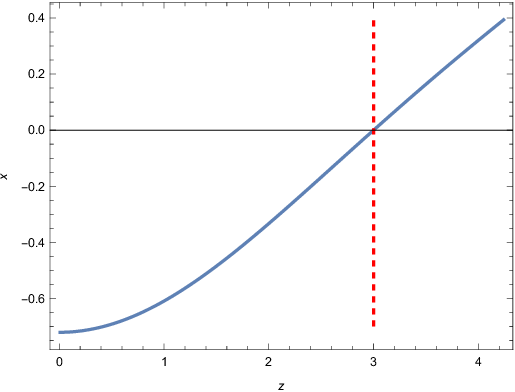}
\includegraphics[width=3in]{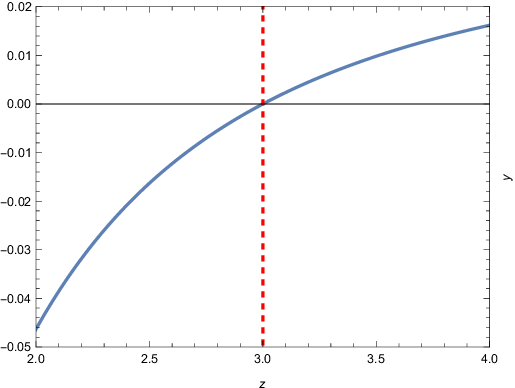}
\end{center}
  \caption{At the onset of the instability, the normalizable component $a_{2,0}$ of the bulk mode $\delta a$
  diverges (the left panel);
  the normalizable component $u_{1,0}$ of the bulk mode $\delta u_a$
  diverges as well (the right panel).
  The vertical dashed red lines
  indicate the critical value of $\kappa$, see \eqref{kcr}.
} \label{figure1}
\end{figure}

The relevant equations are \eqref{a1}-\eqref{a2}.
\nxt In the UV, \ie as $r\to 0_+$, we find
\begin{equation}
\begin{split}
&\delta a=r \biggl(\colorbox{red}{1} \ln r
+\colorbox{yellow}{$a_{2,0}$}\biggr)+r^3 \biggl(
\frac14+\frac{\kappa^2}{36} ( a_{2,0}+1)+\frac14 a_{2,0}-\frac{1}{27} u_{1,0}+\left(\frac14+\frac{\kappa^2}{36}
\right) \ln r\biggr)\\
&+\calo(r^4)\,,
\\
&\delta u_a=\colorbox{yellow}{$u_{1,0}$} r+\biggl(
\frac34 \kappa^2+u_{1,0}-\frac32 \kappa^2 a_{2,0}-\frac32 \kappa^2\ \ln r\biggr) r^2+\calo(r^3\ln r)\,,
\end{split}
\eqlabel{uvh}
\end{equation}
where we highlighted the source term $\colorbox{red}{1}$ (which can be set to 1 as the equations are linear), and the
normalizable coefficients $\colorbox{yellow}{$a_{2,0}$}$ and $\colorbox{yellow}{$u_{1,0}$}$.
\nxt Regularity at the black hole horizon identifies the asymptotic expansions, as  $y\equiv 1-r\to 0_+$, as
\begin{equation}
\begin{split}
&\delta a=a_0^h+\frac{6 a_0^h \kappa^2+27 a_0^h-4 u_0^h}{3(\kappa^2-18)}\ y+\calo(y^2)\,,\\
&\delta u_a=u_0^h+\frac{\kappa^2 (6 a_0^h \kappa^2+27 a_0^h-4 u_0^h)}{2(\kappa^2-18)}\ y+\calo(y^2)\,.
\end{split}
\eqlabel{irh}
\end{equation}
We use the shooting method code developed in \cite{Aharony:2007vg} to
determine $\{a_{2,0}\,, a^h_0\,, u_{1,0}\,, u^h_0\}$ as we vary $\kappa$.
The results of the numerics are presented in fig.~\ref{figure1}.
Both normalizable coefficients $\frac {1}{a_{2,0}}$ and $\frac {1}{u_{1,0}}$ vanish at
\begin{equation}
\kappa_{crit}= 3\,,
\eqlabel{critmu}
\end{equation}
corresponding to $\frac T\mu$ in \eqref{kcr}.

\bibliographystyle{JHEP}
\bibliography{n4rn}

\end{document}